\documentclass[a4paper, 10pt, conference]{ieeeconf}      % Use this line for a4
\usepackage{amsfonts}
\usepackage{amssymb}
\usepackage{amsmath}                                                        % paper
\usepackage{subfig}
\usepackage{epsfig}
\usepackage{graphicx}
\IEEEoverridecommandlockouts                              % This command is only
\newtheorem{theorem}{Theorem}

\title{\LARGE \bf
Minimax Linear Quadratic Gaussian Control of Nonlinear MIMO System with Time Varying Uncertainties}

\author{Obaid Ur Rehman, Ian R. Petersen and  Bar\i\c{s} Fidan% <-this % stops a space
\thanks{Dr. Obaid Ur Rehman is with school of Engineering and Information Technology, University of New South Wales at the Australian Defence Force Academy, Canberra, Australia. ({\tt\small s.obaid.rehman@gmail.com})}%
\thanks{Prof. Ian R. Petersen is with school of Engineering and Information Technology, University of New South Wales at the Australian Defence Force Academy, Canberra, Australia. ({\tt\small i.r.petersen@gmail.com})}%
\thanks{B. Fidan is an assistant professor at Mechanical and Mechatronics Engineering Department, University of Waterloo, N2L3G1 ON, Canada
{\tt\small fidan@uwaterloo.ca}.}}

\begin{document}

\maketitle
\thispagestyle{empty}
\pagestyle{empty}

%%%%%%%%%%%%%%%%%%%%%%%%%%%%%%%%%%%%%%%%%%%%%%%%%%%%%%%%%%%%%%%%%%%%%%%%%%%%%%%%
\begin{abstract}

In this paper, a robust nonlinear control scheme is proposed for a nonlinear multi-input multi-output (MIMO) system subject to bounded time varying uncertainty which satisfies a certain integral quadratic constraint condition. The scheme develops a robust feedback linarization approach which uses standard feedback linearization approach to linearize the nominal nonlinear dynamics of the uncertain nonlinear system and linearizes the nonlinear time varying uncertainties at an arbitrary point using the mean value theorem. This approach transforms uncertain nonlinear MIMO systems into an equivalent MIMO linear uncertain system model with unstructured uncertainty. Finally, a robust minimax linear quadratic Gaussian (LQG) control design is proposed for the linearized model. The scheme guarantees the internal stability of the closed loop system and provides robust performance. In order to illustrate the effectiveness of this approach, the proposed method is applied to a tracking control problem for an air-breathing hypersonic flight vehicle (AHFV).
\end{abstract}

%%%%%%%%%%%%%%%%%%%%%%%%%%%%%%%%%%%%%%%%%%%%%%%%%%%%%%%%%%%%%%%%%%%%%%%%%%%%%%%%
\section{INTRODUCTION}

The general problem of feedback linearization for nonlinear systems with uncertainty has been typically approached in the literature by imposing some conditions on the uncertainty description which are known as matching conditions \cite{NBSS} and the strict triangularity condition \cite{FB_robust_Marino}. Methods considering mismatched uncertainties also exist, in which uncertainties are decomposed into matched and mismatched parts. These methods typically require the mismatched parts not to exceed some maximum allowable  bound \cite{FBoutput}. In an attempt to solve a related issue arising in feedback linearization, in our previous work \cite{Rehman_ASJC,Rehman_ASCC01} we propose a method of robust feedback linearization to feedback linearize a nonlinear system with uncertainties by representing the uncertainties in a realistic way and relaxing the matching condition requirements on the description of the uncertainties. In this approach, we linearized the nominal part of the system using the feedback linearization approach and linearized the remaining nonlinear terms with respect to each uncertainty and state to obtain an acceptable linear form for the uncertainty model at arbitrary operating points. However, in order to express the system in a more convenient set of  coordinates, we have defined a diffeomorphism $T$ which depends on the nominal values (without uncertainty) of the parameters. This definition of the diffeomorphism requires that the system either satisfies the generalized matching conditions \cite{NBSS} which are relaxed versions of the matching condition or allows for additional uncertainty inputs in the system. 

In order to relax the generalized matching condition and the strict triangularity requirement, in this paper, we have introduced a notion of an uncertain diffeomorphism. This definition of uncertain diffeomorphism is similar to the one used in our previous work (see \cite{CDC02}). Furthermore, in order to deal with the nonlinear uncertain terms which are subject to time varying uncertainty, here we use a mean value approach similar to the approach used in \cite{Rehman_ASCC01}. The uncertain diffeomorphism used in this paper is the function of system states and uncertain time varying parameters. Generally, feedback linearization of higher order systems involves higher order derivatives of the system's outputs being required to be measurable. However, in the real world applications, it is not possible to measure or manipulate all of the output derivatives (new states) algebraically especially in the presence of uncertain diffeomorphism. The minimax LQG control approach solves this problem by using output feedback and estimating the unmeasured states in the presence of the uncertainties. The main idea in our approach is to suppress the perturbations arising from the nonlinear uncertainties using a minimax LQG controller \cite{IP}. 

In the later part of this paper, we apply the proposed method to design a tracking controller for velocity and altitude tracking of an air-breathing hypersonic flight vehicle (AHFV) in the presence of input coupling and flexible mode effects. In this paper, we solve the AHFV tracking control problem by designing a robust minimax LQG controller in combination with the robust feedback linearized model proposed in this paper. In the minimax LQG scheme, the uncertain states are estimated by using a robust Kalman filter. 

The main contributions of this work as compared to previously published work \cite{Rehman_ASJC, Rehman_ASCC01, CDC02} are as follows:
\begin{itemize}
\item [1.] Feedback linearization of uncertain systems subject to \textbf{time varying uncertainty} using an uncertain diffeomorphism along with a \textbf{mean value approach}.
\item [2.] Estimation of uncertain states and design of a tracking controller using the minimax LQG design method for the linearized model.
\end{itemize}

The paper is organized as follows. Section \ref{sec:system} describes the class of nonlinear systems and uncertainties considered in the paper. A complete derivation of the robust feedback linearization of the uncertain system is presented in Section \ref{sec:FB}. In Section \ref{sec:MMX}, the minimax LQG control design method is presented for the feedback linearized system with linearized uncertainty. For the case study using an uncertain nonlinear model of the AHFV, the uncertainty modeling and control design methods with tracking simulation results  are presented in Section \ref{sec:example}. Conclusions are presented in Section \ref{sec:disc}.

\section{System Definition}\label{sec:system}
Here, we consider an uncertain multi-input multi-output nonlinear system having same number of inputs and outputs and which is subject to time varying uncertainty $p(t)$:
\begin{small}
\begin{equation}
\label{eqsystem}
\begin{split}
\dot{x}(t)&=f(x(t),p(t))+\sum\limits_{k=1}^m{g_k(x(t),p(t))u_k(t)},\\
y_i(t)&=\nu_i(x(t)),\quad i=1,2,\cdots,m
\end{split}
\end{equation}
\end{small}
where $x(t)\in \mathbb{R}^n$, $u(t)=[u_1.....u_m]^T\in\mathbb{R}^m$ and $y(t)=[y_1....y_m]^T\in\mathbb{R}^m$. The nonlinear functions $f(x(t),p(t))$, and $g_k(x(t),p(t))$ and $\nu_i(x(t),p(t))$ for $i=1,\cdots,m$ are infinitely differentiable (or differentiable to a sufficiently large degree) functions of their arguments. Also, $p(t)\in \mathbb{R}^q \in \Omega$ is a vector of unknown parameters or disturbances which takes values in the set $\Omega\subset \mathbb{R}^p$. The subscript indices $k$ and $i$ indicate $kth$ and $ith$ elements of the corresponding vectors respectively. The full state vector $x(t)$ is assumed to be available for measurement and the uncertainty in the system satisfies an integral quadratic constraint condition (IQC) (see \cite{IP}). It is assumed that the uncertain functions  can be written as $f(x(t),p(t)) = f(x(t),p_0)+\Delta f(x(t),p(t))$ and  $g(x(t),p(t)) = g(x(t),p_0)+\Delta g(x(t),p(t))$ where $p_0$ is the nominal value of the parameter. In addition, the uncertain functions $\Delta f(x(t),p(t))$ and $\Delta g(x(t),p(t))$ are smooth and contain all the uncertainties in the system, including disturbances and uncertain nonlinear terms. Furthermore, there exist an isolated equilibrium point which is not affected by the vector $p(t)$; i.e. $f(0)=0$,  and $\Delta f(0,p(t))=0$ and system has full relative degree with respect to the regulated output.

\section{Robust Feedback Linearization}\label{sec:FB}
In this section a robust feedback linearized method is used to linearize (\ref{eqsystem}) using a technique developed in our previous work; see \cite{Rehman_ASJC,Rehman_ASCC01}. We decompose the system (\ref{eqsystem}) into nominal and uncertain parts as follows:
\begin{scriptsize}
\begin{equation}
\label{eqSSsystem}
\begin{split}
\dot{x}(t)&=\underbrace{f_0(x(t),p_0)+\sum\limits_{k=1}^m{g_{k0}(x(t),p_0)u_k(t)}}_\text{Nominal part}\\
&+\underbrace{\Delta f(x(t),p(t))+\sum\limits_{k=1}^m{\Delta g_k(x(t),p(t))u_k(t)}}_\text{Uncertain part},\\
y_i(t)&=\nu_i(x(t)),\quad i=1,2,\cdots,m.
\end{split}
\end{equation}
\end{scriptsize}
The nominal nonlinearities in the equation (\ref{eqsystem}) can be canceled using a standard feedback linearization approach \cite{IS}.  Let us assume that $r_i$ for $i=1,2,\cdots,m$ is the relative degree of each regulated output, the Lie derivative of each output $\nu_i$, $r_i$ number of times, for each subsystem can be written as follows (we drop the argument $t$ from the functions for the sake of brevity):
\begin{footnotesize}
\begin{equation}
\label{eqidiffoutput}
\begin{array}{c}
y_i^1=L_{f_0}(\nu_i)+L_{\Delta f}(\nu_i),\\
y_i^2=L_{f_0}^{2}(\nu_i)+L_{\Delta f}^{2}(\nu_i),\\
\vdots\\
y_i^{r_1}=L_{f_0}^{r_1}(\nu_i)+\sum\limits_{k=1}^m L^{r_1-1}_{g_k}[L_{f_0}( \nu_i)]u_k+L_{\Delta f}^{r_1}(\nu_i)\\
+\sum\limits_{k=1}^m L^{r_1-1}_{ \Delta g_k}[L_{ \Delta f}( \nu_i)]u_k.
\end{array}
\end{equation}
\end{footnotesize}
In order to write the system in a form suitable for feedback linearization, we write the $r_i^{th}$ derivative of each output as follows:
\begin{scriptsize}
\begin{align}
\label{eqdiffoutput}
\left[\begin{array}{c}
y_1^{r_1}\\
\vdots \\
y_m^{r_m}
\end{array}\right]
&=f_*(x,p_0)+g_*(x,p_0)u\nonumber\\
&+\left[\begin{array}{c}
L_{\Delta f}^{r_1}(\nu_1)+\sum\limits_{k=1}^m L^{r_1-1}_{ \Delta g_k}[L_{ \Delta f}( \nu_1)]u_k\\
\vdots\\
L_{\Delta f}^{r_m}(\nu_m)+\sum\limits_{k=1}^m L^{r_m-1}_{\Delta g_k}[L_{\Delta f}(\nu_m)]u_k
\end{array}\right],
\end{align}
\end{scriptsize}
where,
\begin{scriptsize}
\begin{align*}
f_*(x,p_0)&=[L_{f_0}^{r_1}(\nu_1) \cdots L_{f_0}^{r_m}(\nu_m)]^T,\nonumber\\
g_*(x,p_0)&=\left[
\begin{array}{ccc}
L_{g_{10}}L_{f_0}^{r_1-1}(\nu_1)& \dots& L_{g_{m0}}L_{f_0}^{r_1-1}(\nu_1)\\
L_{g_{10}}L_{f_0}^{r_2-1}(\nu_2)&\dots &L_{g_{m0}}L_{f_0}^{r_2-1}(\nu_2)\\
\quad\vdots \quad&\quad&\vdots \\
L_{g_{10}}L_{f_0}^{r_m-1}(\nu_m)&\dots& L_{g_{m0}}L_{f_0}^{r_m-1}(\nu_m)
\end{array}\right],
\end{align*}
\end{scriptsize}
and the Lie derivative of the functions $\nu_i$ with respect to the vector fields $f$ and $g_k$ are given by
\begin{scriptsize}
\begin{align*}
L_{f} \nu_i=&\frac{\partial \nu_i (x)}{\partial x} f ,~ L_{f}^j\nu_i=L_{f} (L_{f}^{j-1}\nu_i (x))
,~ L_{g_k }(\nu_i)=\frac{\partial \nu_i (x)}{\partial x}g_k.
\end{align*}
\end{scriptsize}
The nominal feedback linearizing control law
\begin{small}
\begin{equation}
\label{equ}
u=-g_*(x,p_0)^{-1}f_*(x,p_0)+g_*(x,p_0)^{-1}v
\end{equation}
\end{small}
partially linearizes the input-output map (\ref{eqdiffoutput}) in the presence of uncertainties as follows:
\begin{small}
\begin{align}
\label{eqsdiffoutputm}
&y^{r_i}_*=\underbrace{\left[
\begin{array}{c}
v_1\\
\vdots\\
v_m
\end{array}\right]}_\text{Nominal part}+
\underbrace{\left[
\begin{array}{c}
\Delta W_1^{r_1}(x,u,p_0,\Delta p(t))\\
\vdots\\
\Delta W_m^{r_m}(x,u,p_0,\Delta p(t))
\end{array}\right]}_\text{Uncertainty part},
\end{align}
\end{small}
where
$\Delta W_i^{r_i}(x,p_0,\Delta p(t))=L_{\Delta \bar{f}}^{r_i}( \nu_i)+\sum\limits_{k=1}^m L^{r_i-1}_{\Delta g_k}[L_{\bar{f}}( \nu_i)]u_k$,  $y_*=[y_1^{r_1} .... y_m^{r_m}]^T$, and $v=[v_1 .... v_m]^T$ is the new control input vector. Furthermore, we define an uncertainty vector \begin{small}$\Delta W_i$\end{small} which represents the uncertainty in each derivative of the $i^{th}$ regulated output as
\begin{small}
\begin{equation}
\label{eqWi}
\begin{split}
\Delta W_i(x,u,p_0,\Delta p(t))&=\left[\begin{array}{c}
\Delta W_i^{1}(x,u,p_0,\Delta p(t))\\
\Delta W_i^{2}(x,u,p_0,\Delta p(t))\\
\vdots\\
\Delta W_i^{r_i}(x,u,p_0,\Delta p(t))
\end{array}\right]\\
&=
\left[\begin{array}{c}
L_{\Delta \bar{f}}(\nu_i)\\
L_{\Delta \bar{f}}^{2}(\nu_i)\\
\vdots\\
L_{\Delta \bar{f}}^{r_i}( \nu_i)+\sum\limits_{k=1}^m L^{r_i-1}_{\Delta g_k}[L_{\bar{f}}( \nu_i)]u_k
\end{array}\right],
\end{split}
\end{equation}
\end{small}
and write $y_i$ for $i=1,2,\cdots,m$ as given below.
\begin{small}
\begin{equation}
\label{eqyi}
\left[\begin{array}{c}
y_i^1\\
y_i^2\\
\vdots\\
y_i^{r_i}
\end{array}\right]=\left[\begin{array}{c}
0\\
0\\
\vdots\\
v_i
\end{array}\right]+\left[\begin{array}{c}
\Delta W_i^{1}(x,u,p_0,\Delta p(t))\\
\Delta W_i^{2}(x,u,p_0,\Delta p(t))\\
\vdots\\
\Delta W_i^{r_i}(x,u,p_0,\Delta p(t))
\end{array}\right].
\end{equation}
\end{small}
Let us define an uncertain diffeomorphism for each partially linearized system in (\ref{eqyi}) for $i=1,\cdots, m$ as given below:
\begin{equation}
\label{eqdiffm}
\begin{split}
\chi_i&=T_i(x,p(t))=\\
&\left[\begin{array}{ccccc}
\int y_i-y_{ic} & y_i-y_{ic} & y^{1}_i & .. & y^{r_i-1}_i\end{array}\right]^{T}.
\end{split}
\end{equation}
Using the diffeomorphism (\ref{eqdiffm}) and system (\ref{eqyi}) we obtain the following: 
\begin{equation}
\label{eqfullBvsky}
\dot{\chi}=A{\chi}+B v+\Delta \bar{W}(\chi,v,p_0,\Delta p(t)),\quad\quad
\end{equation}
where $\chi(t)=[\chi_1(t),\cdots,\chi_m(t)]^T\in \mathbb{R}^{\bar{n}}$, $v(t)=[v_1~v_2 \cdots v_m]^T\in\mathbb{R}^m$ is the new control input vector, $ \Delta \bar{W}(\chi,v,p(t))=[\Delta \bar{W}_1(\cdot), \Delta \bar{W}_2(\cdot),~ \cdots ~,  \Delta \bar{W}_m]^T$ is a transformed version of $\Delta W (x,u,p(t))$ and $\Delta\bar{W}_i(\cdot)=[0, 0,~ \cdots,~ w_i^{(r_i)}(\cdot)]^T$ for $i=1,2,~\cdots,~m$. Also,
\[
A=\left[
\begin{array}{ccc}
A_1&  \dots  & 0\\
\vdots   & \ddots & \vdots \\
0 & \dots & A_m
\end{array}\right];
~~B=\left[
\begin{array}{cccc}
\bar{B}_1&  \dots  & 0\\
\vdots   & \ddots & \vdots \\
0 &  \dots & \bar{B}_m
\end{array}\right].
\]
In order to obtain a fully linearized form for (\ref{eqfullBvsky}), here, we use a similar approach as used in \cite{Rehman_ASCC01}. In this work, we perform the linearization of $\Delta \bar{W}(\chi,v,p)$ using the generalized mean value theorem \cite{Lin_algebra, vector_valued_mean} such that no higher order uncertain terms exist after the linearization process. Since in this scheme an uncertain diffeomorphism is used, therefore this scheme provides a bound which would be less conservative than the bound obtained in \cite{Rehman_ASCC01}.
%$\chi=0\in \mathfrak{B}$ and $v=0\in\mathfrak{B}$
\begin{theorem}\cite{vector_valued_mean} \label{th_mean}
Let $\bar{w}_i^{(j)}:\mathbb{R}^n \rightarrow \mathbb{R}$ be differentiable on $\mathbb{R}^n$ with a Lipschitz continuous gradient $\nabla \bar{w}_i^{(j)}$. Then for given $\chi$ and $\chi(0)$ in $\mathbb{R}^n$, there is a $c=\chi+\bar{t}(\chi-\chi(0))$ with $\bar{t}\in [0,1]$, such that
\begin{equation}
\label{eqvect_mean}
\bar{w}_i^{(j)}(\chi)-\bar{w}_i^{(j)}(\chi(0)))=\nabla \bar{w}_i^{(j)}(c_i).(\chi-\chi(0)).
\end{equation}
\end{theorem}
In order to extend Theorem \ref{th_mean} to the case of $\bar{w}:\mathbb{R}^n \rightarrow \mathbb{R}^{\bar{m}}$ we can write
\begin{equation*}
\bar{w}(\chi)-\bar{w}(\chi(0))=\bar{w}'(c).(\chi-\chi(0)),
\end{equation*}
where $\bar{w}'$ is the Jacobian of the function $\bar{w}(\chi)$ and $c$ is a point on the straight line between $\chi$ and $\chi(0)$ which may be different for different rows of $\bar{w}'(c)$ \cite{vector_valued_mean}. We may estimate the norm of $\bar{w}(\chi)-\bar{w}(\chi(0))$ as follows:
\begin{equation}
\label{eq_mean_norm}
\begin{split}
\Vert \bar{w}(\chi)-\bar{w}(\chi(0))) \Vert &=\Vert \bar{w}'(c).(\chi-\chi(0)) \Vert\\
&\leq \Vert \bar{w}'(c) \Vert \Vert (\chi-\chi(0)) \Vert.
\end{split}
\end{equation}
The Lipschitz constant $\bar{w}$ may be estimated by $max_c\Vert\bar{w}'(c)\Vert$  where, $\Vert \cdot \Vert$ represents the Euclidean norm.

We can apply the result of Theorem \ref{th_mean} to the nonlinear uncertain part of (\ref{eqfullBvsky}). Let us define a hyper rectangle
\begin{equation}
\mathfrak{B}=\{\left[
\begin{array}{c}
\chi\\
v
 \end{array}\right]:\begin{array}{c}
\underline{\chi_i}\leq\chi_i\leq\bar{\chi_i}\\
\underline{v_i}\leq v_i\leq\bar{v_i}
 \end{array}\},
\end{equation}
where $\underline{\chi_i}$, and $\underline{v_i}$  denote the lower bounds and $\bar{\chi_i}$, and $\bar{v_i}$ denote the upper bounds on the new states and inputs respectively. For this purpose, the Jacobian of $w_i^{(j)}(\cdot)$ is found by differentiating it with respect to $\chi$ and $v$ at an arbitrary operating point $c_{ij}=[\tilde{\chi} ~ \tilde{v} ~ \tilde{p}(t)]$ for $i=1,2,\cdots,m$ and $j=1,2,\cdots, r_i$ where, $\tilde{\chi}$,  $\tilde{v}\in\mathfrak{B}$, and $\tilde{p}(t)\in\Omega$. Since we assume $w_i^{(j)}(0,0,p(t))=0 $, $\chi(0)=0$, and $v(0)=0$; $w_i^{(j)}(\cdot)$ can be written as follows:
\begin{equation}
w_i^{(j)}(\chi,v,\Delta p(t))  = {w'}_i^{(j)}(c_{ij})\cdot[\chi \quad v \quad \Delta p(t)]^T.
\end{equation}
And then $\Delta \bar{W}(\cdot)$ can be written as 
\begin{small}
\begin{equation}
\label{eqbarW}
\Delta\bar{W}(\cdot)= \Phi \left[
\begin{array}{c}
\chi\\
v\\
\Delta p(t)
 \end{array}\right],
\end{equation}
\end{small}
where,
\begin{small}
\[
\Phi=\left[
\begin{array}{c}
 {w'}_{1}^{(r_1)}(c_{1r_1}) \\
\vdots\\
 {w'}_{2}^{(r_2)}(c_{2r_2})\\
\vdots\\
 {w'}_{m}^{(r_m)}(c_{m r_m}) 
\end{array}\right].
\]
\end{small}
Also, the bound on $\Delta \bar{W}(\cdot)$ can be obtained as follows:
\begin{equation}
\label{eqboundm}
\tilde{\rho}=\max_{c_{ir_i}} \Vert \Phi \Vert,\quad   i=1,2,\cdots,m.
\end{equation}
The bound in (\ref{eqboundm}) is obtained by over bounding $\Phi(t)$.
\subsection{Linearized model with an Unstructured Uncertainty Representation}
In (\ref{eqbarW}), $c_{(\cdot)}$ is chosen such that it gives the maximum induced matrix norm on $\Phi$. Once these bounds are obtained, we can write (\ref{eqfullBvsky}) in a suitable MIMO stochastic uncertain system form so that the minimax LQG control approach \cite{IP} can be utilized to design a tracking controller. We define, $\zeta_1(t)=\Delta_1(t)[\bar{C}_1 \chi(t)+\bar{D}_1 v(t)]\in \mathbb{R}^{m}$,  and $\bar{W}(t)=[\zeta(t) \quad \tilde{w}_2^T]^T\in \mathbb{R}^{m+n}$, where $\zeta= \zeta_1(t) + \tilde{w}_1$, and $\tilde{w}_1$ is a disturbance input corresponding to $\Delta p(t)$. Also $\tilde{w}_2$ is a unity covariance noise input. We write linearized model as follows:
\begin{equation}
\label{eqgform1}
\begin{split}
\dot {\chi}(t) &=A\chi(t)+B_1 v(t)+ B_2 \bar{W}(t);\\
z(t)&=C_1{\chi(t)}+D_1 v(t);\\
\tilde{y}(t)&=C_2{\chi(t)}+D_2 \bar{W}(t);\\
\end{split}
\end{equation}
where,
\[
B_1=B,~B_2=\left[
\begin{array}{cc}
B_1 E_1 & \mathbf{0}
\end{array}\right],~~
\bar{C}_1=\left[
\begin{array}{cccc}
0 & \cdots & \tilde{\rho} & 0 \\
0 & 0 & \cdots & \tilde{\rho}
\end{array}\right]
\]
\[
\bar{D}_1=\tilde{\rho}\mathbf{I}_2,\quad D_2=\left[
\begin{array}{cc}
\mathbf{0} & I 
\end{array}\right],\quad E_1=\mathbf{E}[\Delta p(t) \Delta p(t)^T],
\]
$C_1=E_1^{-1}\bar{C}_1$, $D_1=E_1^{-1}\bar{D}_1$ and $\Vert\Delta\Vert\leq 1$. Note that $\Delta(t)$ is $m \times m$ which satisfies the following stochastic uncertainty constraint condition.
\begin{scriptsize}
\begin{equation}
\label{eqIQC}
\mathbf{E}\int_0^{\infty}\Vert {\zeta_1} \Vert^2 \leq \mathbf{E}\int_0^{\infty}\Vert {z} \Vert^2,
\end{equation}
\end{scriptsize}
where $\Vert . \Vert$ indicates the Euclidean norm.. 

\section{Minimax LQG Design}\label{sec:MMX}
The model developed in above section uses an uncertain diffeomorphism $T(x,p(t))$ which is unknown and hence any control system design using this model must contains a robust filter which able to estimates the uncertain states. Therefore, in this section we propose a minimax LQG design approach which uses a robust Kalman filter to estimates the elements of $T(x,p(t))$ and guarantees the stability and robust performance of the closed loop system. Here we present a summary of the minimax LQG design procedure. Interested readers are referred to \cite{IP} for more details on results and related proofs.

The minimax LQG control problem \cite{IP} involves  finding a controller which minimizes the maximum value of the following cost function:
\begin{small}
\begin{equation}
\label{eqfcost}
J=\lim_{T\rightarrow\infty} \frac{1}{2T}\mathbf{E}\int_0^{T}(\chi(t)^T R \chi(t)+v(t)^T G v(t))dt,
\end{equation}
\end{small}
where $R>0$ and $G>0$.
The maximum value of the cost is taken over all uncertainties satisfying the uncertainty constraint (\ref{eqIQC}).
If we define a variable
\begin{small}
\begin{equation}
\label{eqPsi}
{\Psi}=\left[\begin{array}{c}
R^{1/2} \chi\\
G^{1/2} v
\end{array}\right],
\end{equation}
\end{small}
the cost function (\ref{eqfcost}) can be written as follows:
\begin{small}
\begin{equation}
\label{eqffcost}
J=\lim_{T\rightarrow\infty} (\frac{1}{2T})\mathbf{E}\int_0^{T}\Vert \Psi \Vert^2 dt.
\end{equation}
\end{small}
The minimax optimal controller problem can now be solved by solving a scaled risk-sensitive control problem \cite{IP} which corresponds to a scaled $H_\infty$ control problem; e.g. see \cite{IP_LQG}. In this control problem the system is described by (\ref{eqgform1}) and (\ref{eqPsi}) and the controller is to be constructed such that the closed loop system is stable and the transfer function  from $W(t)$ to $\Psi$ satisfies the $H_{\infty}$ norm bound $\Vert T_{W\Psi}(jw)\Vert\leq 1 \forall w$. The scaled risk-sensitive control problem considered here allows a tractable solution in terms of the following pair of $H_{\infty}$ type algebraic Riccati equations.
\begin{scriptsize}
\begin{align}
\label{eqARE1}
&(A-B_2 D_2^T  \Gamma^{-1} C_2)Y_{\infty}+Y_{\infty}(A-B_2 D_2^T \Gamma^{-1}  C2)^T \nonumber\\
&- Y_{\infty}(C_2^T \Gamma^{-1}C_2-\tau^{-1}R_{\tau})Y_{\infty} + B_2(I-D_2^T  \Gamma^{-1} D_2)B_2^T =0,
\end{align}
\end{scriptsize}
and
\begin{scriptsize}
\begin{align}
\label{eqARE2}
&X_{\infty}(A-B_1 G_{\tau}^{-1}  \Upsilon_{\tau}^T )+(A-B_1 G_{\tau}^{-1} \Upsilon_{\tau}^T )^T  X_{\infty}\nonumber \\
&- X_{\infty}(B_1 G_{\tau}^{-1}B_1^T -\tau^{-1}B_2 B_2^T )X_{\infty} + (R_{\tau}-\Upsilon_{\tau} G_{\tau}^{-1}\Upsilon_{\tau}^T )=0,
\end{align}
\end{scriptsize}
where,
\begin{scriptsize}
\[
R_{\tau}\triangleq R+\tau C_1^T  C_1,\quad G_{\tau} \triangleq G+\tau D_1^T  D_1,\quad \gamma_{\tau}\triangleq \tau C_1^T  D_1, \quad \Upsilon_{\tau}\triangleq D_1D_1^T.
\]
\end{scriptsize}
The solutions to both of the algebraic Riccati equations are required to satisfy the conditions $Y_{\infty} > 0$, $X_{\infty}>0$, $I-\tau^{-1} Y_{\infty} X_\infty > 0$ and $R_\tau-\Gamma_\tau ^T G_{\tau}^{-1}\Gamma_\tau \geq 0$.
In order to solve the minimax LQG control problem, the parameter $\tau > 0$ is chosen to minimize the cost bound ($W_\tau$) defined by
\begin{small}
\begin{align}
\label{eqbound}
&W_\tau \triangleq tr[(\tau Y C_2^T+B_2 D_2^T)(D_2D_2^T)^{-1}\nonumber\\
&\times (\tau C_2 Y + D_2 B_2^T)X(I-YX)^{-1}+\tau Y R_{\tau}].
\end{align}
\end{small}
The minimax LQG controller $H(s)$ has the following form:
\begin{small}
\begin{align}
\label{eqcontroller}
\dot{\hat{\chi}}&=A_c \hat{\chi} + B_c \tilde{y};\nonumber\\
v&=K \hat{\chi},
\end{align}
\end{small}
where, $\hat{\chi}\in\mathbf{R}^{\hat{n}}$ is the state of the controller and
\begin{small}
\begin{align*}
K&=-G_{\tau}^{-1}(B_1^T  X_{\infty}+\Upsilon_{\tau}^T );\\
B_c&=(I-\tau^{-1} Y_{\infty} X_\infty )^{-1}(Y_{\infty} C_2^T  + B_2 D_2^T )\Gamma^{-1};\\
A_c&=A+B_1 K- B_c C_2 + \tau^{-1}(B_2-B_c D_2)B_2^T  X_{\infty}.
\end{align*}
\end{small}

\section{Air-breathing hypersonic flight vehicle example}\label{sec:example}
\subsection{Vehicle Model}\label{sec:VM}
In this section, we consider the same example as considered in our previous work \cite{Rehman_ASCC01,CDC02}. The nonlinear model for the longitudinal dynamics of an AHFV is presented in \cite{LPV01}:
\begin{small}
\begin{equation*}
\label{eqn1}
\dot {V}=\frac{T \cos \alpha -D}{m}- g \sin\gamma,~~
\dot {\gamma }=\frac{L+T\sin \alpha }{mV}-\frac{g \cos \gamma }{V},
\end{equation*}
\begin{equation}
\label{eqAHFVmodel}
\dot {h}=V\sin \gamma,\quad
\dot {\alpha }=Q-\dot {\gamma },~~
\dot {Q}=M_{yy} /I_{yy},
\end{equation}
\begin{equation*}
\label{eqn3}
\ddot{n_i}=-2\zeta_{m} w_{m,i} \dot{n_i}-w_{m,i}^2 n_i+N_i,\qquad i=1,2,3.
\end{equation*}
\end{small}
Approximations to the forces and moments occurring in these equations are given as follows:
\begin{small}
\begin{equation}
L\approx \bar{q} S C_L (\alpha ,\delta_e,\delta_c,\Delta \tau_1,\Delta \tau_2 ),
\end{equation}
\begin{equation}
D\approx \bar{q} S C_D (\alpha ,\delta_e,\delta_c,\Delta \tau_1,\Delta \tau_2 ),
\end{equation}
\begin{equation}
M_{yy} \approx z_T T+\bar{q} S \bar {c} C_M(\alpha ,\delta_e,\delta_c,\Delta \tau_1,\Delta \tau_2 ),
\end{equation}
\begin{equation}
T\approx \bar{q}  [\phi C_{T,\phi}(\alpha,\Delta \tau_1,M_{\infty})+C_T(\alpha,\Delta \tau_1,M_{\infty},A_d)],
\end{equation}
\begin{equation}
N_i\approx\bar{q} C_{N_i} [\alpha ,\delta_e,\delta_c,\Delta \tau_1,\Delta \tau_2],\quad i=1,2,3.
\end{equation}
\end{small}
The coefficients obtained by fitting curves corresponding to these quantities are given as follows; here, we remove the function arguments for the sake of brevity:
\begin{footnotesize}
\begin{equation}
\label{eqCL}
\begin{split}
C_L&=C_L^\alpha \alpha+C_L^{\delta_e} \delta_e+C_L^{\delta_c} \delta_c+C_L^{\Delta\tau_1}\Delta\tau_1+C_L^{\Delta\tau_2}\Delta\tau_2+C_L^0,\\
C_M&=C_M^\alpha \alpha+C_M^{\delta_e} \delta_e+C_M^{\delta_c} \delta_c+C_M^{\Delta\tau_1}\Delta\tau_1+C_M^{\Delta\tau_2}\Delta\tau_2+C_M^0,\\
C_D &= C_D^{(\alpha+\Delta\tau_1)^2} (\alpha+\Delta\tau_1)^2+C_D^{(\alpha+\Delta\tau_1)} (\alpha+\Delta\tau_1)
+C_D^{\delta_e^2} \delta_e^2 \\
&+C_D^{\delta_e} \delta_e+C_D^{\delta_c^2} \delta_c^2+C_D^{\delta_c} \delta_c
+C_D^{\alpha\delta_e}\alpha\delta_e+C_D^{\alpha\delta_c}\alpha\delta_c \\
&+C_D^{\delta\tau_1} \delta\tau_1+C_D^0,\\
C_{T,\phi}&=C_{T,\phi}^\alpha \alpha+C_{T,\phi}^{\alpha M_{\infty}^{-2}} \alpha M_{\infty}^{-2}+ C_{T,\phi}^{\alpha \Delta\tau_1} \alpha \Delta\tau_1+C_{T,\phi}^{M_{\infty}^{-2}} M_{\infty}^{-2} \\
&+C_{T,\phi}^{{\Delta\tau_1}^2} {\Delta\tau_1}^2+C_{T,\phi}^{{\Delta\tau_1}}{\Delta\tau_1}+C_{T,\phi}^0,\\
C_T&=C_T^{A_d}A_d+C_T^\alpha \alpha+C_{T}^{M_{\infty}^{-2}} M_{\infty}^{-2}+C_{T}^{{\Delta\tau_1}}{\Delta\tau_1}+C_T^0,\\
C_{N_i}&=C_{N_i}^\alpha \alpha+C_{N_i}^{\delta_e} \delta_e+C_{N_i}^{\delta_c} \delta_c+C_{N_i}^{\Delta\tau_1}\Delta\tau_1+C_{N_i}^{\Delta\tau_2}\Delta\tau_2+C_{N_i}^0,
\end{split}
\end{equation}
\end{footnotesize}
where $n=[n_1~ n_2~ n_3]^T$, and $E_j\in \mathbb{R}^{1\times 3}$ are vectors which describe the linear relationship $\Delta \tau_j=E_j n$ for $j=1,2$ \cite{LPV01}. The terms $M_{\infty}$ and $\bar{q}$ are defined as follows:
\begin{equation}
\bar{q}=\frac{\rho(h) V^2}{2},\quad
M_{\infty}=\frac{V}{M_0}.
\end{equation}
The nonlinear equations of motion in (\ref{eqAHFVmodel}) have $11$ states (in the vector $x$) in which there are five rigid body states; i.e., velocity $V$, altitude $h$, angle of attack $\alpha$, flight path angle $\gamma$, and pitch rate $Q$ and there are three vibrational modes which are represented by generalized modal coordinates $n_i$, where $N_i$ is a generalized force. There are four inputs (in the vector $u$) and they are the diffuser-area-ratio $A_d$, the throttle setting or fuel equivalence ratio $\phi$, the elevator deflection ($\delta_e$), and the canard deflection $\delta_c$. 
For tracking control purposes we simplify the model is such a way that the scheme developed in Section \ref{sec:FB} can be used and the simplified model closely approximates the real model (see also, \cite{Rehman_MSC01,Rehman_GNC01}). Note that the effect of structural flexibility is entering into the system (\ref{eqAHFVmodel}) through the forebody turn angle and aftbody vertex angle, $\Delta \tau_1$ and $\Delta \tau_2$ of the vehicle respectively. In the process of simplification, firstly, we remove all the flexible states $n_j$ for $j=1,2$ from the CFM and consider the effect of flexibility in the model by considering $\Delta \tau_1(x)$ and $\Delta \tau_2(x)$ as uncertain parameters. We simplify the forces and moment coefficients as follows:
\begin{small}
\begin{align}
\label{simp_coeff}
&C_{T,\phi}=C_{T,\phi}^\alpha \alpha+C_{T,\phi}^{\alpha M_{\infty}^{-2}} \alpha M_{\infty}^{-2}\alpha+C_{T,\phi}^{M_{\infty}^{-2}}  M_{\infty}^{-2}+C_{T,\phi}^0\nonumber\\
&+\Delta C_{T,\phi}(x,u),\nonumber\\
&C_L =C_L^\alpha \alpha+C_L^0+\Delta C_l(u),\nonumber\\
&C_M=C_M^\alpha \alpha+[C_M^{\delta_e}-C_M^{\delta_c}(\frac{C_L^{\delta_e}}{C_L^{\delta_c}})] \delta_e+C_M^0+\Delta C_M(x,u),\nonumber\\
&C_D = C_D^{(\alpha+\Delta\tau_1)^2} (\alpha)^2+C_D^{(\alpha+\Delta\tau_1)} (\alpha)+C_D^0+\Delta C_d(x,u),\nonumber\\
&C_T=C_T^{A_d}A_d+C_T^\alpha \alpha+C_{T}^{M_{\infty}^{-2}} M_{\infty}^{-2}+C_T^0+\Delta C_T(x,u),
\end{align}
\end{small}
where \begin{small}$\Delta C_d(\cdot)$, $\Delta C_l(\cdot)$, $\Delta C_T(\cdot)$, $\Delta C_{T,\phi}(\cdot)$ and $\Delta C_M(\cdot)$\end{small} represent the uncertainties in their corresponding functions. Furthermore, in order to obtain full relative degree for the purpose of feedback linearization, we dynamically extend the system by introducing second order actuator dynamics (adding two more states $\phi$ and $\dot{\phi}$) into the fuel equivalence ratio input as follows:
\begin{small}
\begin{equation}
\label{eqfuel}
\ddot {\phi }=-2\zeta \omega _n \dot {\phi }-\omega _n^2 \phi +\omega
_n^2 \phi _c.
\end{equation}
\end{small}
After this extension, the sum of the elements of vector relative degree will be equal to the order of the system $n$; i.e. $n=7$ and thus satisfying one of the conditions for exact feedback linearization \cite{IS}. 

\subsection{Robust Feedback Linearization of the Simplified Model}\label{sec:FBUlqr}
The model obtained through the above simplification is still difficult to feedback linearize due to the presence of uncertainties in the system. We approach this problem by using the technique developed in Section \ref{sec:FB}. The outputs to be regulated are selected as the velocity $V$ and the altitude $h$ using two inputs, elevator deflection $\delta_e$ and fuel equivalence ratio $\phi_c$. Since $\delta_c$ is a function of $\delta_e$; i.e related to $\delta_e$ via an interconnect gain, we do not consider it as a separate input. Furthermore, we fix the diffuser area ratio $A_d$ to be unity. This manipulation results in a $2$-input and $2$-output square system. The new simplified model consists of seven rigid states and two additional integral states as follows:

\begin{equation}
x=\left[\begin{array}{ccccccccccc}
V_I & V & h_I & h & \gamma & \alpha & \phi & \dot{\phi} & Q
\end{array}\right]^T,
\end{equation}
where,
\[
V_I=\int_0^{t}{(V(\tau)-V_c)}d\tau ,\quad
h_I=\int_0^{t}{ (h(\tau)-h_c)}d\tau,
\]
and $V_c$ and $h_c$ are the desired command values for the velocity and altitude respectively.
The uncertain parameter vector $p\in \mathbb{R}^{9}$ includes the vehicle inertial parameters, coupling terms and the coefficients which appear in the force and moment approximations described previously and is given as follows:
\begin{align}
\label{equpm}
p=[
&C_L^\alpha \quad C_M^{\delta_c} \quad C_{T,\phi}^{\alpha M_{\infty}^{-2}} \quad C_{T,\phi}^{M_{\infty}^{-2}} \Delta C_l \quad \Delta C_d \quad \Delta C_T \quad \nonumber \\
&\Delta C_M \quad \Delta C_{T,\phi}]^T \in \mathbb{R}^{9}.
\end{align}
The model of the AHFV can be written in the form (\ref{eqSSsystem}) as follows:
\begin{align}
\label{equsystemlqr}
\dot{x}(t)&=\bar{f}(x,p_0)+\sum\limits_{k=1}^2{g_k(x,p_0)u_k} +\Delta \bar{f}(x,p)\nonumber\\
&+\sum\limits_{k=1}^2{\Delta g_k(x,p)u_k};\nonumber \\
y_i(t)&=\nu_i(x,p), \quad i=1,2
\end{align}
where, $\Delta \bar{f}(x,p)$ and $\Delta g_k(x,p)$ are the uncertainty terms appearing in the corresponding functions. The control vector $u$ and output vector $y$ are defined as
\[
u=[u_1, u_2]^T=\left[ {\delta _e, \phi _c } \right]^T, y=[y_1, y_2]^T=\left[ {V,h}
\right]^T.
\]
We assume that $p(t) \in \Omega$, where $\Omega$ is a compact convex set that represents the admissible range of variation of $p(t)$ such that $p_0$ lies in its interior. In this study, a maximum variation of $10\%$ of the nominal values has been considered. Thus, $\Omega=\{p(t)\in \mathbb{R}^{9}~ |~ |0.9 p_{i0}| \leq |p_i(t)| \leq |1.1 p_{i0}|~~ \text{for}\quad i=1,\cdots,9\}$.
It is worth mentioning the fact that there are no uncertainty terms exists in $\dot{V}$, and $\dot{h}$, we can write linearized input-output map for the original model (\ref{eqAHFVmodel}) using (\ref{eqsdiffoutputm}) as follows:
\begin{footnotesize}
\begin{equation}
\label{equyfblin}
\left[\begin{array}{c}
\dot{V}\\
\ddot{V}\\
\stackrel{\dots}{V} \\
\dot{h}\\
\ddot{h}\\
\stackrel{\dots}{h} \\
h^{4}
\end{array}\right]=\left[\begin{array}{c}
0\\
0\\
v_1 \\
0\\
0\\
0\\
v_2
\end{array}\right]+
\left[\begin{array}{c}
0\\
\Delta\ddot{V}\\
\Delta\stackrel{\dots}{V}(x,u,p) \\
0\\
\Delta\ddot{h}\\
\Delta\stackrel{\dots}{h}(x,u,p) \\
\Delta {h}^4(x,u,p)\end{array}\right].
\end{equation}
\end{footnotesize}
Corresponding uncertain diffeomorphisms for each system as in (\ref{eqdiffm}) which maps the new vectors $\xi$ and $\eta$ respectively to the original vector $x$ can be written as follows:
\begin{small}
\begin{equation}
\xi=T_1(x,p(t),V_c)
,\quad
\eta=T_2(x,p(t),h_c),
\end{equation}
\end{small}
where,
\begin{small}
\begin{align*}
&T_1(x,p(t),V_c)=\left[
\begin{array}{cccc}
\int_0^{t}{(V(\tau)-V_c)}d\tau & V-V_c & \dot{V} & \ddot{V}\end{array}\right]^T,\\
&T_2(x,p(t),h_c)=\left[
\begin{array}{ccccc}
\int_0^{t}{ (h(\tau)-h_c)}d\tau & h-h_c & \dot{h} & \ddot{h} & \stackrel{\dots}{h}\end{array}\right],
\end{align*}
\end{small}and $V_c$ and $h_c$ are the desired command values for the velocity and altitude respectively.
Also, 
\begin{footnotesize}
\begin{equation}
\label{eqtransformx}
\chi = T(x,p(t),V_c,h_c),
\end{equation}
\end{footnotesize}
where \begin{small} $\chi= \left[\begin{array}{ccccccccc}
\xi_1 & \xi_2 & \xi_3 &\xi_4 & \eta_1& \eta_2& \eta_3& \eta_4& \eta_5 \end{array}\right]^T$\end{small}, and
\begin{footnotesize}
$T(x,p(t),V_c,h_c)=\left[\begin{array}{cc}
T_1(x,p(t),V_c) & T_2(x,p(t),h_c) \end{array}\right]^T$
\end{footnotesize}.
And finally we can rewrite (\ref{equyfblin}) using the method given in Section \ref{sec:FB} as follows:
\begin{footnotesize}
\begin{align}
\label{eqlfinal}
\left[\begin{array}{c}
\dot{\xi}_1 \\
\dot{\xi}_2 \\
\dot{\xi}_3 \\
\dot{\xi}_4 \\
\dot{\eta_1}\\
\dot{\eta_2}\\
\dot{\eta_3}\\
\dot{\eta_4}\\
\dot{\eta}_5
\end{array}\right]&=\left[\begin{array}{c}
0\\
0\\
0\\
v_1 \\
0\\
0\\
0\\
v_2
\end{array}\right]
+
\left[\begin{array}{c}
0\\
0\\
0 \\
\Delta w_{1}(\tilde{\chi},\tilde{v},p(t)) \\
0\\
0\\
0 \\
0 \\
\Delta w_{2}(\tilde{\chi},\tilde{v},p(t))
\end{array}\right]\chi\\
&+
\left[\begin{array}{c}
0\\
0\\
0\\
\Delta \tilde{w}_{1}(\tilde{\chi},\tilde{v},p(t)) \\
0\\
0\\
0\\
0\\
\Delta \tilde{w}_{2}(\tilde{\chi},\tilde{v},p(t))
\end{array}\right]v.
\end{align}
\end{footnotesize}
Furthermore, we  can rewrite (\ref{eqlfinal}) in the general form (\ref{eqgform1}) where,
\begin{equation*}
\begin{split}
A&=\left[\begin{array}{ccccccccc}
0 & 1 & 0 & 0 & 0 & 0 & 0 & 0 & 0\\
0 & 0 & 1 & 0 & 0 & 0 & 0 & 0 & 0\\
0 & 0 & 0 & 1 & 0 & 0 & 0 & 0 & 0\\
0 & 0 & 0 & 0 & 0 & 0 & 0 & 0 & 0\\
0 & 0 & 0 & 0 & 0 & 1 & 0 & 0 & 0\\
0 & 0 & 0 & 0 & 0 & 0 & 1 & 0 & 0\\
0 & 0 & 0 & 0 & 0 & 0 & 0 & 1 & 0\\
0 & 0 & 0 & 0 & 0 & 0 & 0 & 0 & 1\\
0 & 0 & 0 & 0 & 0 & 0 & 0 & 0 & 0\\
\end{array}\right],~~
B=\left[\begin{array}{cc}
0 & 0  \\
0 & 0  \\
0& 0  \\
1 & 0  \\
0 & 0  \\
0 & 0  \\
0 & 0 \\
0 & 0  \\
0 & 1 \\
\end{array}\right];
\\
C_1&=\left[
\begin{array}{ccccccccc}
0& 0 & 0 & 10 \times \tilde{\rho} & 0 & 0 & 0 & 0 & 0\\
0 & 0 & 0 & 0 & 0 & 0 & 0 & 0 & 10 \times \tilde{\rho}\\
\end{array}\right],\\
D_1&=\left[
\begin{array}{cc}
10 \times\tilde{\rho} & 0  \\
0 & 10 \times\tilde{\rho} \\
\end{array}\right],\quad
B_2 =[0.1 \times B\quad \mathbf{0}_6],
\end{split}
\end{equation*}
$\Vert\Delta(t)\Vert\leq 1$ and  $\chi(t)\in \mathbb{R}^9$ is the state vector, $v(t)=[v_1~v_2]^T\in\mathbb{R}^2$ is the new control input vector and $\zeta(t)=\Delta[C_1 \chi(t)+D_1 v(t)]\in \mathbb{R}^{2}$ is the uncertainty output. It is worth noting that the states  $\xi_1,~ \xi_2,~ \xi_3,~\eta_3,~\eta_4,~\eta_5$ can either be measured or constructed with the available hardware on board the aircraft.

\begin{figure*}[h]
\begin{center}
\epsfig{file=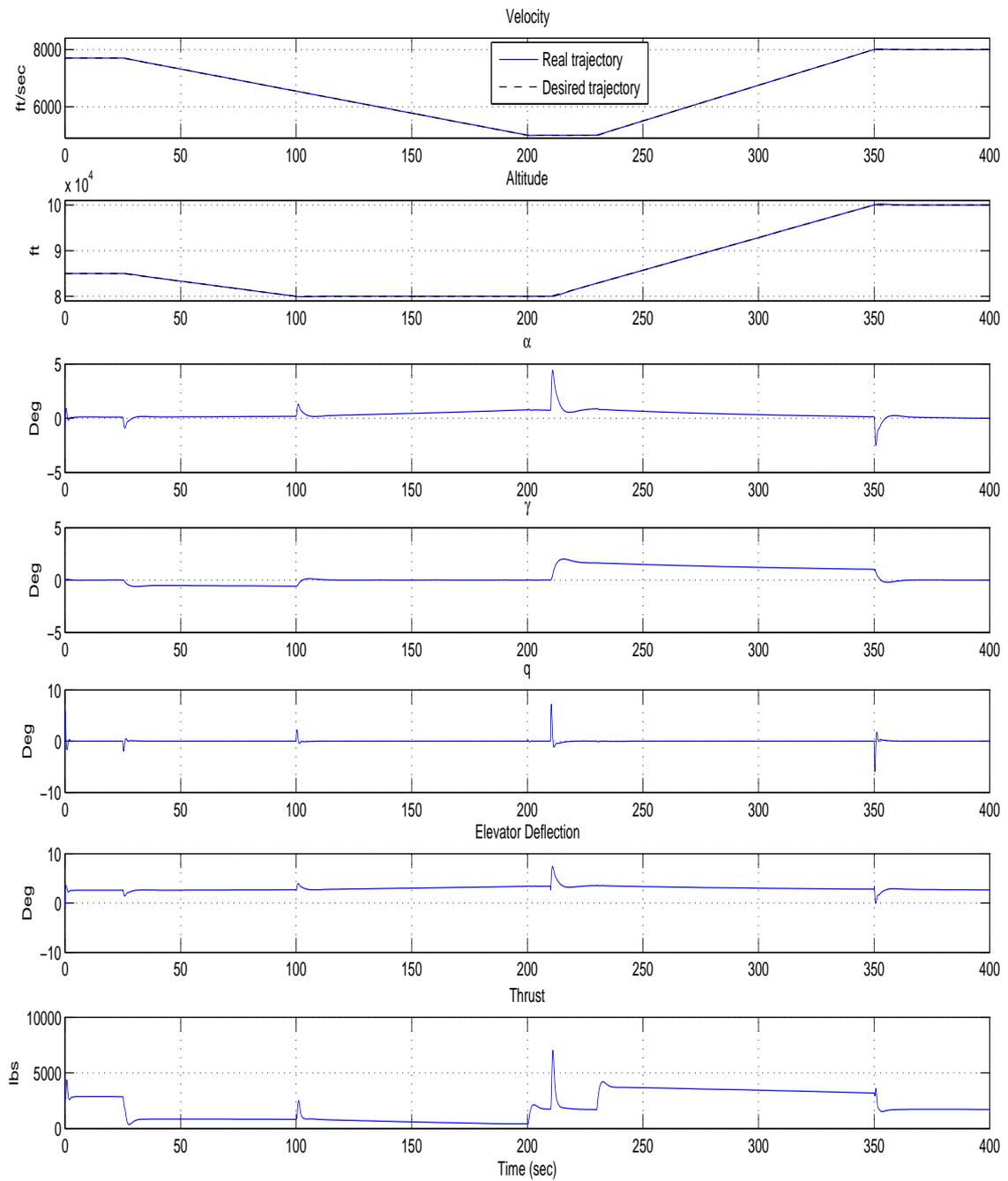, height=8in,width=7.0in}
\caption{Velocity and altitude reference tracking responses using the mean value approach.}
\label{fig:vel_alt_traj_res_mv}
\end{center}
\end{figure*}

Finally, we design a minimax LQG controller as described in Section \ref{sec:MMX} for AHFV velocity and altitude tracking control problem. The simulation result of the proposed controller with the original CFM (without simplification) in the presence of time varying uncertainty is shown in Fig. \ref{fig:vel_alt_traj_res_mv}. In the simulation certain velocity and altitude trajectories have been chosen to demonstrate the effectiveness of the approach. It is observed that the velocity and altitude tracking is achieved while control input and other states remain bounded.

\section{Conclusion} \label{sec:disc}
A robust nonlinear control scheme for an uncertain nonlinear system with time varying uncertainty is presented using robust feedback linearization and minimax linear quadratic Gaussian (LQG) methods. In the proposed method, a linearized uncertainty model is derived for the corresponding uncertain nonlinear system which is followed by a minimax LQG controller. The applicability of the scheme to a real world application is demonstrated by designing a robust tracking controller for an air-breathing hypersonic flight vehicle model with input coupling and flexible effects. The approach involves the linearization of a simplified curve fitted model using a robust feedback linearization method as the first step. In the second step, a velocity and altitude tracking controller is synthesized using the minimax LQG control design method. Simulation results with a large flight envelope simulation is also presented to demonstrate the effectiveness of the scheme. The results show that the proposed method works very well under parameter uncertainties and give satisfactory results. Further, investigation of the results reveals that the minimax LQG based controller works well with parameter variations of up to $10$\% of their nominal values for which it is designed.
%%%%%%%%%%%%%%%%%%%%%%%%%%%%%%%%%%%%%%%%%%%%%%%%%%%%%%%%%%%%%%%%%%%%%%%%%%%%%%%%
\section{ACKNOWLEDGMENTS}
This research was supported by the Australian Research Council.

%%%%%%%%%%%%%%%%%%%%%%%%%%%%%%%%%%%%%%%%%%%%%%%%%%%%%%%%%%%%%%%%%%%%%%%%%%%%%%%%

%References are important to the reader; therefore, each citation must be complete and correct. If at all possible, references should be commonly available publications.

\bibliographystyle{IEEEtran}        % Include this if you use bibtex
\bibliography{Literature_jabref}   

\end{document}